\documentclass[graybox]{svmult}

\usepackage{mathptmx}       
\usepackage{helvet}         
\usepackage{courier}        
\usepackage{type1cm}        
\usepackage{makeidx}         
\usepackage{graphicx}        
\usepackage{multicol}        
\usepackage[bottom]{footmisc}

\usepackage{amssymb}
\usepackage{amsmath}

\newcommand{\td}{\mathrm{d}}

\newcommand{\sun}{\odot}

\usepackage[square,sort,comma,numbers]{natbib}
\makeindex             

\begin{document}

\title*{Stellar evolution and modelling stars}
\author{V\'ictor Silva Aguirre}
\institute{V\'ictor Silva Aguirre \at Stellar Astrophysics Centre, Department of Physics and Astronomy, Aarhus University, Ny Munkegade 120, DK-8000 Aarhus C, Denmark, \email{victor@phys.au.dk}}
%
%
\maketitle

%
\abstract{In this chapter I give an overall description of the structure and evolution of stars of different masses, and review the main ingredients included in state-of-the-art calculations aiming at reproducing observational features. I give particular emphasis to processes where large uncertainties still exist as they have strong impact on stellar properties derived from large compilations of tracks and isochrones, and are therefore of fundamental importance in many fields of astrophysics.}
\section{Stars, why bother?}\label{sec:intro}
A comprehensive view of how stars evolve and interact with their surrounding medium is crucial for many fields in astrophysics. Stars are the main sources of chemical evolution in the Universe (e.g., \cite{2016JPhCS.703a2004M}), and their physical, chemical, and kinematic characteristics preserve information about their birth environment and subsequent evolution to present age. For these reasons, the integrated properties of stellar populations can be used as tools to understand the evolution of distant galaxies (see e.g., \cite{Conselice:2014ct} and references therein). When individual stars can be resolved by observations, as is the case in our Galaxy, accurately characterising them can help unveiling the main processes responsible for the formation history and evolution of the Milky Way (see \cite{BlandHawthorn:2016iq} for a review).

Stars are also the progenitors of objects and events of high astrophysical importance, such as supernova, gamma ray burst, planetary nebulae, and stellar black holes (e.g., \cite{Maoz:2014gd,Smartt:2009kr,Kumar:2015kk,Herwig:2005jn,2007coaw.book.....C} and references therein). Last, and definitely not least, stars are the hosts of exoplanets, and our ability to characterise planetary systems (from size and mass to their atmospheric composition) depends critically on having an accurate representation of the parent star's properties (e.g., \cite{Winn:2015jt}). You only know your planet as well as you know your star. From these examples it is clear that stars are the building blocks of astrophysics, and understanding them holds the key to unveil the relevant physical processes in many areas of astronomical research. We must discover what are the main drivers of stellar evolution and the impact in observable properties that can help us decoding the wonderful nature of stars.

The luminosity flux radiated from the surface of a star is a natural consequence of the intricate processes taking place in the stellar interior. How this flux evolves and changes with time, the field concerning with stellar structure and evolution, is routinely described to a considerable extent as a `solved' problem in astrophysics. However, and although the main ingredients of the theory of stellar evolution are well established, large uncertainties still remain in cases where empirical evidence is lacking or the physical descriptions are not well constrained and thus parametrised to a simpler form. How stars produce, transport, and emit their energy; how they burn and mix chemical elements; how is the stellar matter constituted and what is its state, etc., are some of the many relevant topics where still many uncertainties exist.

There are several excellent textbooks on stellar structure and evolution written by notable astronomers, and I encourage the reader to explore titles such as \cite{1968pss..book.....C} (recently updated by \cite{Weiss:2005vv}), \cite{Clayton:1983tb}, \cite{Hansen:2004ul}, and \cite{2012sse..book.....K}, just to name a few. Based on these references I will introduce the basic concepts of stellar structure and evolution emphasising those relevant to asteroseismic studies, and refer the reader when necessary to textbooks where the relevant (and interesting!) details can be found.
\section{Modelling stars}\label{sec:mod}
Stars owe their brightness to a delicate interplay between gravitational contraction and thermonuclear reactions as energy generation sources, both mechanisms not always acting simultaneously. This interaction produces changes in the star on different time scales, which become relevant at different stages of stellar evolution.

When nuclear reactions are not efficient in producing energy, the physical conditions in stellar interiors evolve with the rate of change of the gravitational potential and internal energy. We can consider that the star contracts gradually while maintaining sphericity and that this is the sole responsible for the stellar luminosity. The corresponding time scale a star can shine by this mechanism is called the \textit{Kelvin-Helmholtz} time scale, 
\begin{equation}\label{sec:mod,eq:kh_tim}
\tau_{\mathrm{kh}} \simeq \frac{GM^2}{2RL}\,,
\end{equation}
where $G$ is the Newtonian constant of gravitation, $M$ is the mass of the star, $R$ its radius, and $L$ the associated luminosity produced by the contraction. For a 1~M$_\sun$ star, this is approximately 15~Myr.

Stellar matter is naturally victim of interacting forces such as gravitation and pressure, but if no acceleration of material takes place a stage of mechanical equilibrium is maintained. This equilibrium in a star, referred to as hydrostatic equilibrium, is one of the pillars of stellar structure studies. The shortest relevant time scale describes the time a star needs to recover its equilibrium when the balance between gravitational forces and pressure is disturbed by some dynamical process, for instance a pressure wave. If a star is close to hydrostatic equilibrium, it corresponds to the free-fall time scale of the star, or the \textit{dynamical} time scale
\begin{equation}\label{sec:mod,eq:dyn_tim}
\tau_{\mathrm{dyn}} \simeq \sqrt{\frac{R^3}{GM}} \simeq \sqrt{\frac{1}{G \, \langle\rho\rangle}}\,,
\end{equation}
where $\langle\rho\rangle$ corresponds to the mean density of the star. In the case of the Sun, the free-fall time scale is of the order of 20~min.

Finally, the longest relevant time scale involved is the \textit{nuclear} time scale,
\begin{equation}\label{sec:mod,eq:nuc_tim}
\tau_{\mathrm{nuc}} = \frac{\epsilon qMc^2}{L}\,,
\end{equation}
where $c$ is the speed of light, $q$ the fraction of the total stellar mass involved in the nuclear burning, and $\epsilon$ the amount of mass that is converted into energy as a result of the nuclear reaction processes. Essentially, this time scale describes how long a star can shine with nuclear fusion as its sole source of energy. If the Sun was made of pure hydrogen and the central 10\% of its mass would contribute to nuclear reactions, it could shine through this mechanism for approximately 10~Gyr.

There are large differences between the relevant time scales for stellar evolution, where $\tau_{\mathrm{nuc}}\gg \tau_{\mathrm{kh}} \gg \tau_{\mathrm{dyn}}$. If we intent to follow a large portion of the evolution of a star, the processes occurring in the shortest time scales must be consistently parametrised over larger periods. Moreover, some processes are usually neglected due to the lack of a consistent theory describing their effects in the overall stellar evolution, such as rotation and magnetic fields. Therefore, many assumptions and simplifications about the nature of stellar matter and complex physics must be used when modelling stars.
\subsection{The main equations}\label{ssec:eqs}
The vast majority of stars are currently in long-lasting phases of their evolution, in which the time scales involved for appreciable change to occur are too large to be observed. Thus, the evolutionary changes are described by the four basic differential equations of stellar structure which are dubbed the macrophysics in modern theory of stellar studies:
\begin{eqnarray}
\frac{\partial r}{\partial m}  = \frac{1}{4\pi r^2 \rho}\,, \label{ssec:eqs,eq:hyd} \\
\frac{\partial P}{\partial m}  = -\frac{Gm}{4\pi r^4}\,, \label{ssec:eqs,eq:mass} \\
\frac{\partial L}{\partial m}  = \epsilon-\epsilon_\nu + \epsilon_g\,, \label{ssec:eqs,eq:cons} \\
\frac{\partial T}{\partial m}  = -\frac{GmT}{4\pi r^4 P} \nabla\,, \label{ssec:eqs,eq:energ}
\end{eqnarray}
where
\begin{equation}
\nabla=\frac{\partial \ln T}{\partial \ln P}\,. \label{ssec:eqs,eq:nabla}
\end{equation}
In these equations, $r$ is the distance from the centre of the star, and $m$ the mass contained within this distance. $P$, $T$ and $\rho$ are the thermodynamic variables pressure, temperature and density respectively, while $L$ is the luminosity at the corresponding position of $r$ (or $m$). The $\epsilon$ term corresponds to the energy rate per unit mass generated by nuclear reactions, $\epsilon_\nu$ to the energy rate lost (in form of neutrinos), and $\epsilon_g$ to the work that is performed on the gas during any expansion or contraction of the star. These equations are, in order of appearance, the \textit{mass conservation equation}, the \textit{hydrostatic equilibrium equation}, the \textit{energy conservation equation}, and the \textit{energy transport equation}.

The solutions to the equations are not stationary but rather evolve with time as a consequence of contraction and nuclear reactions taking place, and the resulting changes in the chemical composition and mean molecular weight brought about by them. The first two equations define the mass profile in the stellar interior, while the latter two equations determine the thermal profile inside of the star. In fact, Eq.~\ref{ssec:eqs,eq:nabla} is simply the definition of $\nabla$, whose value must be derived from a theory of energy transport.

Under normal circumstances there is a steady flow of energy from the deep stellar interior, where the nuclear reactions take place, to the outermost layers of the star, where energy is radiated to the interstellar medium. Depending on the thermodynamical properties of matter in the stellar interior, this energy transport can occur via \textit{radiative transfer}, \textit{convective motions}, or \textit{conductive transfer}. The latter transport method becomes very efficient under degenerate matter conditions.
\subsubsection{Energy transport by radiation}\label{sssec:rad}
In the time-independent three dimensional case, the equation of radiative transfer can be written as
\begin{equation}
\mu_i\frac{\partial \textbf{I}}{\partial x_i} = - (\kappa_\mathrm{ab} + \kappa_\mathrm{sc})\, \rho \textbf{I} + \kappa_\mathrm{ab} \, \rho B + \kappa_\mathrm{sc} \, \rho J\,, \label{sssec:rad,eq:rad_trns}
\end{equation}
where $\textbf{I}(\textbf{x},\boldsymbol\mu,t)$ is the specific intensity at $\textbf{x}$ in direction $\boldsymbol\mu$, $\kappa_\mathrm{ab}$ the mean absorption opacity, $\kappa_\mathrm{sc}$ the scattering opacity,  $B=(ac/4\pi)T^4$ is the integrated Planck intensity and $J$ the mean intensity. In order to estimate the flux carried put by radiation, it is customary to use the \textit{Eddington Approximation} (\cite[e.g.][]{Unno:1966ur}). Assuming the intensity to be isotropic, a relation between the zeroth and first order moments of Eq.~\ref{sssec:rad,eq:rad_trns} can be obtained, that yields the radiation flux as
\begin{equation}
\textbf{F} = - \frac{4 \pi}{3\rho\,(\kappa_\mathrm{ab} + \kappa_\mathrm{sc})} \nabla J\,. \label{sssec:rad,eq:rad_flux}
\end{equation}
The near-isotropy of the radiation intensity is usually associated with a short photon mean free path, where radiation is efficiently trapped (i.e. $\rho\kappa_\mathrm{ab} \rightarrow \infty$). In this case, also referred to as the optically thick case, a diffusive mechanism takes place for radiative energy transport leading to the following expression for the energy flux: 
\begin{equation}
\textbf{F} = - \frac{4 \pi}{3\rho\kappa} \nabla B = - \frac{4acT^3}{3\rho\kappa} \nabla T\,, \label{sssec:rad,eq:rad_flux2}
\end{equation}
where $\kappa=|\kappa_\mathrm{ab} + \kappa_\mathrm{sc}|$. This is known as the \textit{diffusion approximation}.

The radiation flux then clearly depends on the opacities $\kappa$, of which we have so far neglected their natural frequency dependence. This comes from the idea that we can replace the problem of frequency dependence (the \textit{non-grey atmosphere} problem) through some sort of mean opacities. In fact, one can show that a particular average of the opacities can be found by imposing the Eddington approximation in the equation of radiative transfer (Eq.~\ref{sssec:rad,eq:rad_trns}), thus making the problem frequency-independent. These are called \textit{Rosseland} mean opacities, and are defined as
\begin{equation}\label{sssec:rad,eq:ross_mean}
\kappa_\mathrm{ross}^{-1}=\frac{\int_0^\infty \! \frac{1}{\kappa_\nu} \frac{\partial B_\nu}{\partial T}\, \td \nu}{\int_0^\infty \! \frac{\partial B_\nu}{\partial T}\, \td \nu} = \frac{\pi}{a c T^3} \int_0^\infty \! \frac{1}{\kappa_\nu} \frac{\partial B_\nu}{\partial T}\, \td \nu\,,
\end{equation}
where $a$ is the radiation constant, and $B_\nu$ is the monochromatic Planck function.

When essentially all energy is transported outwards by photons (condition of radiative equilibrium), it can be shown that the temperature gradient in Eq.~\ref{ssec:eqs,eq:nabla} takes the form
\begin{equation} \label{sssec:rad,eq:nab_rad}
\nabla_\mathrm{rad}=\frac{3}{16\pi a c G}\frac{\kappa L P}{m T^4}\,
\end{equation}
where $\kappa$ is the Rosseland mean opacity of the stellar matter. A very clear description of the radiation theory in stellar interiors and its connection with stellar atmospheres can be found in \cite{Mihalas:1970ts} (recently updated by \cite{2014tsa..book.....H}).
\subsubsection{Energy transport by convection}\label{sssec:conv}
When the temperature gradient indicated in Eq.~\ref{sssec:rad,eq:nab_rad} is too steep, radiation is not able to carry all the energy outwards and convective instabilities set in. A theory of convective transport includes a stability criterion for convection to take place, and a consistent description of how energy is transported outwards by convective motions.

Based on the displacement analysis of random bubbles inside of the star, a criterion for the onset of convective stabilities can be derived. In a nutshell, let us consider a slight temperature fluctuation in a gas element with respect to its surroundings. Assuming that the element remains in pressure equilibrium with the medium, a temperature increase translates into a density decrease if we consider that the stellar matter obeys an ideal gas law. Thus, this lighter bubble will be lifted upwards by the force of buoyancy. The gas element will travel until it becomes unstable due to turbulence and dissolves into the surrounding gas. For adiabatic motions of bubbles, it can be shown that a layer will remain stable if
\begin{equation}\label{sssec:conv,eq:conv}
\nabla_{\mathrm{rad}}<\nabla_{\mathrm{ad}}+\frac{\varphi}{\delta}\nabla_{\mu}\,,
\end{equation}
where $\nabla_{\mathrm{ad}}$ is the temperature gradient introduced in Eq.~\ref{ssec:eqs,eq:nabla} when the displacement of the bubble takes place adiabatically, and
\begin{equation}\label{sssec:conv,eq:partials}
\varphi = \left(\frac{\partial\ln\rho}{\partial\ln\mu}\right)_{\mathrm{P}, T}, \delta = -\left(\frac{\partial\ln\rho}{\partial\ln T}\right)_{\mathrm{P, \mu}}, \nabla_{\mu}= \left(\frac{d\ln\mu}{d\ln P}\right)\,.
\end{equation}
Equation~\ref{sssec:conv,eq:conv} is the  \textit{Ledoux} criterion for convection (\cite{Ledoux:1947ka}), which takes into account variations in the molecular weight $\mu$ to define the boundaries of convective regions. However, in regions of homogeneous composition one has simply the \textit{Schwarzschild} criterion \citep{Schwarzschild:1958jy}, according to which a region remains stable against convection as long as
\begin{equation}
\nabla_{\mathrm{rad}} < \nabla_{\mathrm{ad}}\,. \label{sssec:conv,eq:sch_crit}
\end{equation}
This is by far the most commonly used criterion when modelling stellar evolution. One reason for this is that once convection sets in a given region of the star, the chemical composition gradient is annihilated by convective mixing and Eq.~\ref{sssec:conv,eq:conv} simplifies to Eq.~\ref{sssec:conv,eq:sch_crit}. However, the use of a different criterion can have large effects in the size of the convective regions, with a subsequent impact on i.e., the resulting luminosity, effective temperature, and main sequence lifetime of the star.

In a radiative (dynamically stable) layer, a displaced element is pushed back by buoyancy forces. This interaction imprints a certain momentum in the gas element, which will overshoot from its original position when descending, becoming lighter than its surroundings and thus ascending again. Such oscillations of gas elements could occur in the form of thin needles, and when they take place adiabatically they are characterised by the \textit{Brunt-V\"ais\"al\"a} frequency:
\begin{equation}
N^2 = \frac{g \delta}{H_p}\left( \nabla_{\mathrm{ad}}-\nabla+\frac{\varphi}{\delta}\nabla_{\mu}   \right)  \,, \label{sssec:conv,eq:bv_freq}
\end{equation}
where $H_p$ is the pressure scale height given by
\begin{equation}
{H_p}^{-1} = - \frac{\td \ln P}{\td r}\,. \label{sssec:conv,eq:pscalh}
\end{equation}
For the displacement to be oscillatory the condition for the frequency $N^2>0$ must be fulfilled, which is the case for convectively stable (radiative) regions (cf, Eq.~\ref{sssec:conv,eq:conv}). From this simple analysis of the stability of a fluid element agains local perturbations it transpires that internal gravity waves cannot occur in convective regions. It is interesting to note that, for stars in the subgiant and red giant phase of evolution, $N^2$ reaches very high values in the core of the star due to the strong central condensation and resulting high value of the local gravity (e.g., \cite{ChristensenDalsgaard:2004ie}). Stars in these evolutionary phases are expected to show a rich spectrum of gravity dominated pulsation frequencies, as beautifully confirmed by asteroseismic observations (e.g., \cite{Kjeldsen:2003ek,Bedding:2010ki,Deheuvels:2011fn}). 

Once a region is found by any stability criteria to be convective, the temperature gradient of that zone needs to be defined (Eq.~\ref{ssec:eqs,eq:nabla}). The usual way to do this is using the \textit{mixing-lenght theory} for convection (MLT) in any of its flavours, most commonly the original formulation proposed by \cite{BohmVitense:1958vy}. The critical free parameter involved in the formulation is the so-called mixing-length parameter $\alpha_\mathrm{MLT}=l/H_p$, where $l$ is the distance a bubble will traverse before dissolving into the surrounding medium. Other formulations such as the full spectrum of turbulent eddies by \cite{Canuto:1991bj,Canuto:1992dz} also require the definition of a convective efficiency to properly determine the temperature gradient in convective regions. The most commonly adopted procedure to calibrate this free parameter is using the solar properties, as described in section~\ref{sssec:cal}.
\subsection{Solving the equations}\label{ssec:solv}
Up to now, I have discussed the problem of stellar evolution calculations based on the structure of stars and physical processes taking place inside of them. The correct description of these processes depends critically on the properties of stellar matter, which are termed the microphysics of stellar evolution.

Inspection of the structure equations (Eqs.~\ref{ssec:eqs,eq:hyd} to~\ref{ssec:eqs,eq:energ}) easily reveals that we are trying to solve a problem for five explicitly shown unknowns ($r$, $\rho$, $P$, $L$, and $T$) through a set of four equations. The missing relation is given by the Equation Of State (EOS), which provides one of the thermodynamic quantities in terms of the others (for instance, $\rho=\rho(P,T,\mu)$, where $\mu$ is just an indicator of the general chemical composition). It is customary to refer to the EOS as one of the \textit{constitutive equations} of stellar structure; the other quantities that enter the equations and need to be defined form the set of constitutive equations. These can be written as,
\begin{eqnarray}
\rho & =\rho \,(P,T,\mu)\,, \label{ssec:solv,eq:eos} \\
c_\mathrm{P} & =c_\mathrm{P} \,(P,T,\mu)\,, \label{ssec:solv,eq:sp_heat} \\
\kappa_\nu & =\kappa_\nu \,(P,T,\mu)\,, \label{ssec:solv,eq:opac} \\
r_{jk} & =r_{jk} \,(P,T,\mu)\,, \label{ssec:solv,eq:rates} \\
\epsilon_\nu & =\epsilon_\nu \,(P,T,\mu)\,, \label{ssec:solv,eq:neutr}
\end{eqnarray}
where $c_\mathrm{P}$ is the specific heat at constant pressure, $\kappa_\nu$ the monochromatic opacity of stellar matter (a particular average of it was introduced in Sect.~\ref{sssec:rad}), $r_{jk}$ the thermonuclear reaction rate transforming nuclei $j$ into nuclei $k$, with the corresponding energy generation rate $\epsilon_{jk}$ given by the product of $r_{jk}$ and the energy released when the transformation takes place. Time evolution of a certain chemical species $X_i$ when only nuclear reactions create or destroy it is given by
\begin{equation}
\frac{\partial X_i}{\partial t} = \frac{m_i}{\rho} \left(\sum_j r_{ji} - \sum_k r_{ik}\right)\,, \label{ssec:solv,eq:chem}
\end{equation}
with the constrain that $\sum_i X_i =1$. If exchange of mass occurs between different stellar layers, diffusive processes also affect the evolution of chemical species. 

\subsubsection{Nuclear reactions}\label{sssec:nuc}
Thermonuclearly fuelled reactions in stellar interiors produce energy that is carried out through radiation, convection or conduction (see Sect.~\ref{ssec:eqs}). I briefly mention here the main channels for burning hydrogen into helium, the longest evolutionary phase in the lifetime of a star, and the reaction network to transform helium into heavier elements.

There are two reaction chains that transform four protons into one $^{4}\mathrm{He}$ nucleus, namely the \textit{p-p chain} and the \textit{CNO cycle} (see Table~\ref{sssec:nuc,tab:chain}). Presence of C, N, or O isotopes is necessary for the CNO cycle to begin, and since they are both destroyed and produced during the cycle they act as catalysts for the reactions. The p-p chain and the CNO cycle usually take place simultaneously in a star but with different efficiencies depending on the total stellar mass.
\begin{table}
\caption[Chains for hydrogen burning]{Reaction networks involved in the \textit{p-p chain} and the \textit{CNO cycle}}\label{sssec:nuc,tab:chain}
\centering
\begin{tabular}[!ht]{ c  c  c }
\multicolumn{3}{c}{\large\textbf{\textit{p-p chain}}}\\
pp I & pp II & pp III \\ 
\hline
$^{1}\mathrm{H}$ +  $^{1}\mathrm{H}$ $\rightarrow$  $^{2}\mathrm{D}$ + $\mathrm{e}^{+}$ +$\nu_{e}$ & 
$^{3}\mathrm{He}$ + $^{4}\mathrm{He}$ $\rightarrow$  $^{7}\mathrm{Be}$ + $\gamma$ &
$^{3}\mathrm{He}$ + $^{4}\mathrm{He}$ $\rightarrow$  $^{7}\mathrm{Be}$ + $\gamma$ \\
$^{2}\mathrm{D}$ + $^{1}\mathrm{H}$ $\rightarrow$  $^{3}\mathrm{He}$ + $\gamma$  & 
$^{7}\mathrm{Be}$ + $\mathrm{e}^{-}$ $\rightarrow$  $^{7}\mathrm{Li}$ + $\nu_{e}$ &
$^{7}\mathrm{Be}$ + $^{1}\mathrm{H}$ $\rightarrow$  $^{8}\mathrm{B}$ + $\gamma$\\
$^{3}\mathrm{He}$ + $^{3}\mathrm{He}$ $\rightarrow$  $^{4}\mathrm{He}$ + $^{1}\mathrm{H}$ + $^{1}\mathrm{H}$ &
$^{7}\mathrm{Li}$ + $^{1}\mathrm{H}$ $\rightarrow$  $^{4}\mathrm{He}$ + $^{4}\mathrm{He}$ &
$^{8}\mathrm{B}$ $\rightarrow$  $^{8}\mathrm{Be}$ + $\mathrm{e}^{+}$ +$\nu_{e}$ \\
 &  & $^{8}\mathrm{Be}$ $\rightarrow$  $^{4}\mathrm{He}$ + $^{4}\mathrm{He}$ \\
\hline
\end{tabular}
\begin{tabular}[!ht]{ c  c }
\multicolumn{2}{c}{}\\
\multicolumn{2}{c}{\large\textbf{\textit{CNO cycle}}}\\
CN cycle & NO cycle \\
\hline
$^{12}\mathrm{C}$ +  $^{1}\mathrm{H}$ $\rightarrow$  $^{13}\mathrm{N}$ + $\gamma$ &
$^{15}\mathrm{N}$ +  $^{1}\mathrm{H}$ $\rightarrow$  $^{16}\mathrm{O}$ + $\gamma$ \\
$^{13}\mathrm{N}$ $\rightarrow$  $^{13}\mathrm{C}$ + $\mathrm{e}^{+}$ +$\nu_{e}$ &
$^{16}\mathrm{O}$ + $^{1}\mathrm{H}$ $\rightarrow$  $^{17}\mathrm{F}$ + $\gamma$ \\
$^{13}\mathrm{C}$ + $^{1}\mathrm{H}$ $\rightarrow$  $^{14}\mathrm{N}$ + $\gamma$ &
$^{17}\mathrm{F}$ $\rightarrow$  $^{17}\mathrm{O}$ + $\mathrm{e}^{+}$ +$\nu_{e}$ \\
$^{14}\mathrm{N}$ + $^{1}\mathrm{H}$ $\rightarrow$  $^{15}\mathrm{O}$ + $\gamma$ &
$^{17}\mathrm{O}$ + $^{1}\mathrm{H}$ $\rightarrow$  $^{14}\mathrm{N}$ + $^{4}\mathrm{He}$ \\
$^{15}\mathrm{O}$ $\rightarrow$  $^{15}\mathrm{N}$ + $\mathrm{e}^{+}$ +$\nu_{e}$ & \\
$^{15}\mathrm{N}$ + $^{1}\mathrm{H}$ $\rightarrow$  $^{12}\mathrm{C}$ + $^{4}\mathrm{He}$ & \\
\hline
\end{tabular}
\end{table}

The nuclear energy generation rate ($\epsilon$) of these channels has different temperature sensitivities, meaning that the conditions in the stellar interior will define the efficiency with which each one of them operates. The p-p chain has an average relation of the order of $\epsilon_\mathrm{pp} \propto T^{4}$ at $T\approx15\times10^6$ K, while the CNO cycle has a higher value of $\epsilon_\mathrm{CNO} \propto T^{18}$ at $T\approx20\times10^6$ K. As an example, in the centre of the Sun $T\approx15\times10^6$ K and more than 90\%\, of the energy budget corresponds to the p-p chain. An important consequence of the temperature sensitivities of the nuclear reaction chains is that, if the H-burning process is dominated by the CNO cycle, it will be confined towards the very central regions of the star. This results in a larger energy flux arising from the innermost regions which favours the presence of a convective core.

Following the exhaustion of hydrogen in the centre, stars begin burning helium as soon as the central temperature increases enough to produce the \textit{triple alpha} (3$\alpha$) reaction:
\begin{align*}
{^{4}\mathrm{He}} + {^{4}\mathrm{He}} &\rightarrow  {^{8}\mathrm{Be}} \\
{^{8}\mathrm{Be}} + {^{4}\mathrm{He}} &\rightarrow  {^{12}\mathrm{C}} + \gamma \\
\end{align*}
The temperature sensitivity is quite strong for the 3$\alpha$ reaction: $\epsilon_{\mathrm{3}\alpha} \propto T^{40}$ at $T\approx10^8$ K. Thus, for the same physical reason as in the case of the CNO cycle, stars burning helium via 3$\alpha$ mechanism have extended convective cores. For the sake of completeness, I mention the other important nuclear reactions involved in the helium burning process:
\begin{align*}
{^{12}\mathrm{C}} + {^{4}\mathrm{He}} &\rightarrow  {^{16}\mathrm{O}} + \gamma \\
{^{16}\mathrm{O}} + {^{4}\mathrm{He}} &\rightarrow  {^{20}\mathrm{Ne}} + \gamma \\
\end{align*}

It is clear by looking at these reactions that helium burning transforms $^{4}\mathrm{He}$ particles mainly into $^{12}\mathrm{C}$, $^{16}\mathrm{O}$ and $^{20}\mathrm{Ne}$. These elements have burning processes in more advanced stages of stellar evolution. I refer the reader to the monograph by \cite{Clayton:1983tb} for a detailed description of the above mentioned reactions, and further explanations on the reaction networks of elements heavier than helium.
\subsection{Choosing the ingredients}\label{ssec:ingr}
Based on the descriptions of the previous section, we now turn our attention to the practicalities of modelling a particular star or stellar population. The target(s) in question will have some set of observed properties that we aim at reproducing with our stellar evolution model, for example colour (or effective temperature), magnitude (or gravity), surface chemical composition, and oscillation frequencies, just to name a few. The task is to select the appropriate ingredients in our evolutionary code that will produce the most realistic representation possible of the star, and this requires some assumptions on the relevant physical processes to be included in the modelling procedure. 

We start by defining the microphysics applied in the calculation as described in section~\ref{ssec:solv}. In the case of the Equation Of State, several compilations relevant for stellar calculations are available such as those computed by the OPAL group (\cite{Rogers:1996iv,Rogers:2002cr}), and the FreeEOS (\cite{2003ApJ...588..862C}). These may need to be complemented at the low-temperature regime by other compilations of EOS, such as the MHD EOS (\cite{Hummer:1988kn}) or that of \cite{1995ApJS...99..713S}.

Regarding the opacity of stellar matter, these require different treatments for the radiative and conductive case. Compilations of Rosseland mean radiative opacities (see Eq.~\ref{sssec:rad,eq:ross_mean}) from the OPAL group (\cite{Iglesias:1996dp}) and Opacity Project (OP, \cite{Badnell:2005ef}) are available, normally complemented at low temperatures by molecular opacities from \cite{1994ApJ...437..879A,Ferguson:2005gn}. For the conductive opacities, relevant in the cases of degenerate matter, calculations by \cite{Itoh:1983hj} or the more updated \cite{Cassisi:2007ey} are commonly adopted.

For the nuclear reaction rates, the most frequently used large compilations of cross sections are those by the NACRE collaboration (\cite{Angulo:1999kp}) and the Solar Fusion (\cite{Adelberger:1998iv,Adelberger:2011fp}). We note in passing that the cross sections of two important astrophysical factors ($S_{34}$ and $S_{1,14}$) have been updated by \cite{Marta:2008dq,Costantini:2008it}, and is highly recommended that the chosen set of thermonuclear reactions includes these latest values. Energy losses via neutrino emission (Eq.~\ref{ssec:solv,eq:neutr}) are determined following prescriptions by e.g., \cite{1994ApJ...425..222H} and \cite{1996ApJS..102..411I}.

Once the desired microphysics of the evolutionary calculation is defined, a few more considerations must be made before we can model our targets. First, one must select a criterion for defining the boundaries of convective regions in the stellar interior, which as described in section~\ref{sssec:conv} is normally done using the Ledoux or Schwarzschild criterion. Second, we must define the value of an efficiency parameter describing the gas motions in zones found to be convective to determine the real temperature gradient (Eq.~\ref{ssec:eqs,eq:nabla}). Then, we must define the initial chemical composition used in our stellar modelling exercise based on some knowledge of the surface abundances of the target. In the following I describe how these parameters are normally obtained in evolutionary calculations, and mention some of the additional mixing processes included in some evolutionary calculations depending on type of star considered.

Before closing this section, I note in passing that not all the aforementioned sets of opacities, EOS, nuclear reactions, etc. are be available in all evolutionary codes, and each modeller should carefully ensure that the combination of microphysics is consistent for each calculation (i.e., the used opacities are calculated for the adopted solar mixture).
\subsubsection{Chemical composition}\label{sssec:comp}
In stellar evolution calculations abundances are represented with the letters $X$, $Y$, $Z$, defining the mass fractions of hydrogen, helium, and all elements heavier than helium (`metals'), respectively. These three quantities must be provided to the evolutionary code performing the calculation, and I review in this section the assumptions adopted in their determination. 

First, we need the observed surface abundance of the star we aim at modelling as obtained by e.g., photometry or spectroscopy. In most cases the observed chemical composition is given in terms of logarithmic abundances ratios with respect to the solar value. For the species $i$ and $j$, this is expressed as
\begin{equation}\label{sssec:comp,eq:ab_rat}
\log (i/j) - \log(i/j)_\sun \equiv [i/j]\,,
\end{equation}
where the dependence of the observed abundances on the solar reference value is clearly established. Normally observations provide a measurement of the bulk stellar metallicity in terms of the iron abundance [Fe/H], and to transform it into an estimate of the mass fraction of heavy elements required in evolutionary calculations it is assumed that
\begin{equation}\label{sssec:comp,eq:mass_rat}
\log (Z/X) - \log(Z/X)_\sun \simeq \mathrm{[Fe/H]}\,.
\end{equation}

As we mentioned above, $Z$ in evolutionary calculations includes every element heavier than helium and thus this transformation has two important consequences. First it assumes that iron is by large the most abundant of the heavy elements in a star, which is normally the case unless excess of alpha-elements is present (such as in metal-poor stars, see e.g., \cite{1985ESOC...21....1S}). Second, since a solar reference value $\log(Z/X)_\sun$ must be defined, it implies that the fraction of each element comprising $Z$ is distributed accordingly to the chosen set of solar abundance ratios (or solar `mixture').

The topic of the atmospheric chemical composition of the Sun has been one of hot debate in the past years. The most recent determinations of the solar individual abundances based on realistic 3D hydrodynamical simulations of stellar atmospheres have considerably decreased the total metallicity of the Sun compared to the older predictions based on simpler 1D atmospheric models. However, these new sets of solar abundances are not yet completely accepted by the astrophysics community mainly due to the discrepancy of the predicted internal sounds speed profile with the results of helioseismology (see e.g., \cite{Serenelli:2009ev}), and therefore both the old (\cite{Grevesse:1993vd,Grevesse:1998cy}) and new (\cite{,Asplund:2009eu,Caffau:2011ik}) determinations are widely applied according to personal preference. The reader is referred to \cite{Bahcall:2005kp,Basu:2008fo} for further explanations on the topic and its implications for helioseismology.

In terms of total metallicity determinations the ratio $(Z/X)_\sun$ varies from $0.0245$ to $0.0183$ among the different compilations, which according to Eq~\ref{sssec:comp,eq:mass_rat} implies that the value of [Fe/H] in a stellar model computed with the same $Z$ and $X$ can vary by $\sim$0.13~dex just by the choice of solar mixture. This must be kept in mind when comparing evolutionary calculations with spectroscopic or photometric metallicities, which can quote uncertainties well below this level. 

Equation~\ref{sssec:comp,eq:mass_rat} is not sufficient to estimate the three necessary components $X$, $Y$, and $Z$, of the stellar composition. Moreover, it is not possible to directly measure helium abundances in stars of temperatures lower than $\sim10^4$~K. Thus, a so called `galactic chemical evolution law' is applied (\cite{Peimbert:1976ch}), which basically relates the amount of fresh helium supplied by stars to the interstellar medium relative to their supply of heavy elements:
\begin{equation}\label{sssec:comp,eq:dydz}
\frac{\Delta Y}{\Delta Z} = \frac{Y - Y_\mathrm{ref}}{Z - Z_\mathrm{ref}}\,.
\end{equation}
The existence and value of $\Delta Y / \Delta Z$ relation is still a matter of debate, and it is usually considered to lie around $1.0\leq\Delta Y / \Delta Z\leq3.0$ (\cite[e.g.,][]{Pagel:1998ht,Jimenez:2003il,Casagrande:2007ck}). In Eq.~\ref{sssec:comp,eq:dydz} one must provide a reference point for $Y$ and $Z$; big bang nucleosynthesis values can be considered ($Z_\mathrm{ref}=0$ and $Y_\mathrm{ref}\sim0.2488$, \cite{Steigman:2010gz}), or the initial $Z$ and $Y$ abundances obtained in the Sun from a solar calibration (see Sect.~\ref{sssec:cal} below). In the latter case, the chosen set of today's surface abundances in the Sun plays an important role, although it has been suggested the initial helium abundance of the Sun could be independent of them (\cite{Serenelli:2010gu}). Finally, the third equation comes from the obvious fact that $X+Y+Z=1$.
\subsubsection{Convective efficiency}\label{sssec:cal}
As mentioned in section~\ref{sssec:conv} the value of the convective efficiency (i.e., $\alpha_\mathrm{MLT}$ under the mixing-length formulation) ultimately defines the value of $\nabla$ in a convective region, but unfortunately it cannot be obtained from first principles. Instead it is normally determined via a standard solar calibration, an iterative procedure in which three input parameters are tuned to reproduce the properties of the Sun. This is an optimisation process where the codes vary the initial composition (two parameters, e.g., $Y$ and $Z$) and the convective efficiency to reproduce the solar radius and luminosity, at present solar age, for a chosen surface composition of the Sun. In the case of the mixing-length theory, the resulting value usually ranges between $1.5 < \alpha_\mathrm{MLT} < 2.5$ depending on the input physics and evolutionary code employed. There is a vast amount of literature devoted to the results of solar calibrations where the interested reader can find more details, such as \cite{ChristensenDalsgaard:1996gf,Bahcall:2001fm}, and references therein. 

In recent years, a new possibility has appeared for determining convective efficiency values from 3D simulations of stellar envelopes. These simulations solve the time-dependent hydrodynamic equations of mass, momentum, and energy conservation and are by design free from adjustable parameters such as the $\alpha_\mathrm{MLT}$ (e.g., \cite{Nordlund:2009wq,Kupka:2017ks}, and references therein). Using suitable averages, it is possible to match the atmospheric stratification of the 3D model with the equivalent from one-dimensional calculations and extract a calibrated convective efficiency (e.g., \cite{Trampedach:2014fo,2015A&A...573A..89M}). These results have been recently implemented in standard stellar evolution codes at solar metallicity (\cite{Salaris:2015be,Mosumgaard:2016vl}) and are being extended to other chemical compositions.
\subsubsection{Additional mixing}\label{sssec:mix}
The last layer of complications in the macrophysics of stellar evolution comes from physical processes that are likely to take place in stars and are still poorly understood. Among these we can mention the cases of atomic diffusion of helium and heavier elements, radiative levitation, rotational mixing, the influence of magnetic fields, stellar winds, etc. (\cite[see][for a review]{Pinsonneault:1997kp}). All these processes require the inclusion of some additional free parameter controlling their efficiency as a function of e.g., mass, temperature, luminosity, etc., and are relevant in different regimes of stellar evolution to accurately reproduce the observational results. As an example, the inclusion of atomic diffusion in standard solar models greatly improved the agreement with helioseismic data via inversion techniques (\cite{ChristensenDalsgaard:1993ck}) and is now regarded a necessary ingredient to accurately reproduce the properties of the Sun. However, diffusion as it operates in the solar case requires the inclusion of additional processes to counteract some of its effects in late-type stars via radiative accelerations (e.g., \cite{1998ApJ...504..559T}). Similarly, overshoot from the convective core is necessary to reproduce the shape of the colour-magnitude diagram of open clusters (e.g., \cite{Maeder:1991to}), but despite extensive theoretical studies there is still no firm veredict on the amount of additional mixing required and its dependence on parameters such as mass or metallicity (e.g., \cite{Zahn:1991uz}).

Asteroseismology is beginning to make substantial contributions in understanding this additional mixing processes, and among these we can mention the first measurements of integrated mass-loss in the red giant branch (\cite{Miglio:2012dm}), constraints in the amount of rotational mixing  in subgiant and red giant stars (\cite{Eggenberger:2012ii}), detection of convective cores and constraints on the overshoot efficiency during the core hydrogen- and helium-burning phases (\cite{SilvaAguirre:2013in,Deheuvels:2016ek,Constantino:2015fu}), as well as the inversion of the rotation profile in {\it Kepler} targets (\cite{Deheuvels:2014kz}). It is expected that these results will further our understanding of these fundamental physical processes and allows us to produce even more realistic models of stellar interiors.
\section{Overview of stellar evolution}\label{sec:over}
\begin{figure}[h]
\centering
\sidecaption
\includegraphics[scale=0.65]{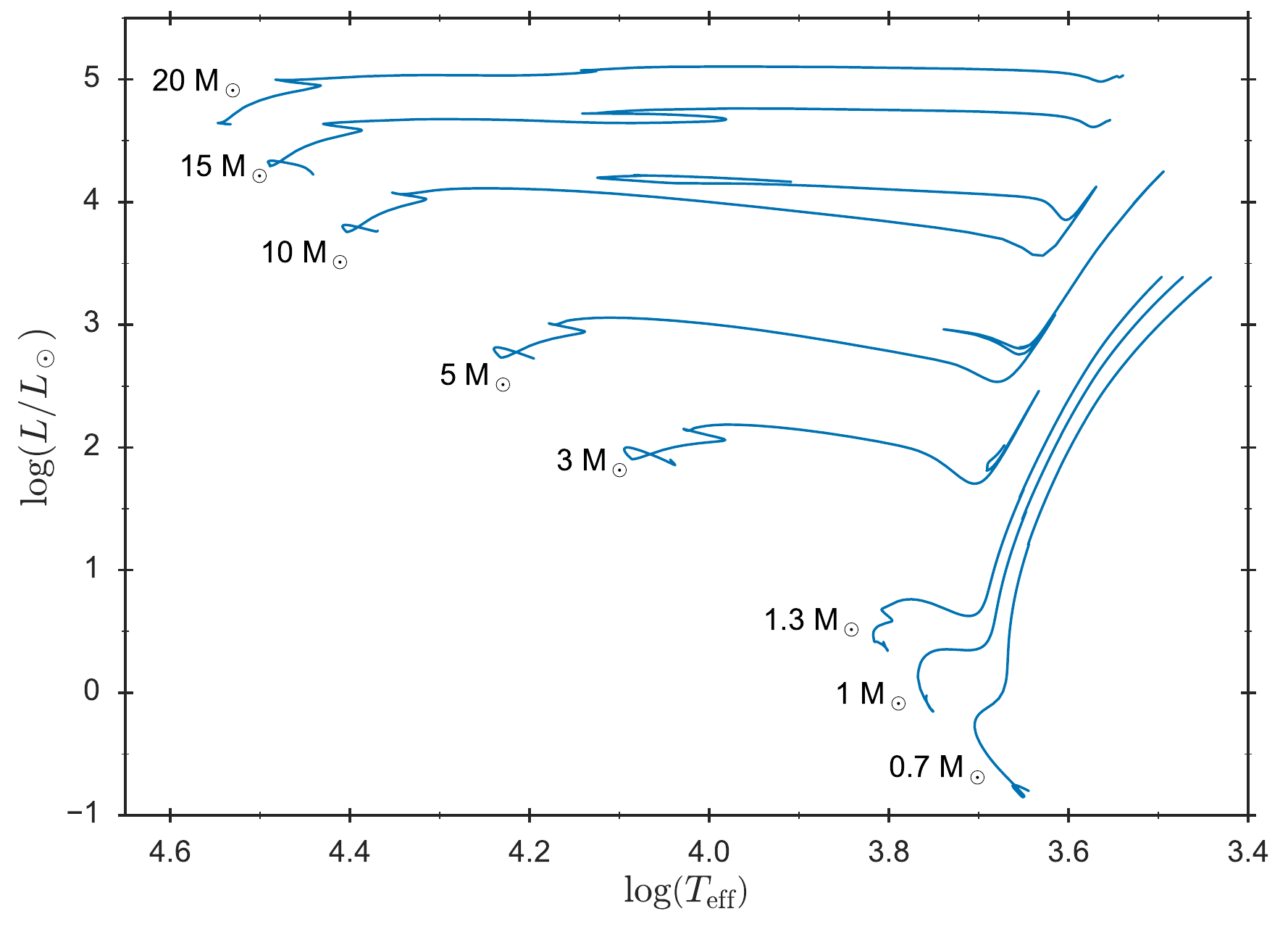}
\caption{Hertzsprung-Russell Diagram plotted for different masses at a fixed chemical composition of $Y=0.28$ and $Z=0.02$. The tracks cover the evolution from the beginning of the main sequence to the red giant phase or, for M~$\geq$~3.0~M$_\sun$, also the helium burning phase.}
\label{fig:hrd}   
\end{figure}

\begin{figure}[h]
\centering
\sidecaption
\includegraphics[scale=0.65]{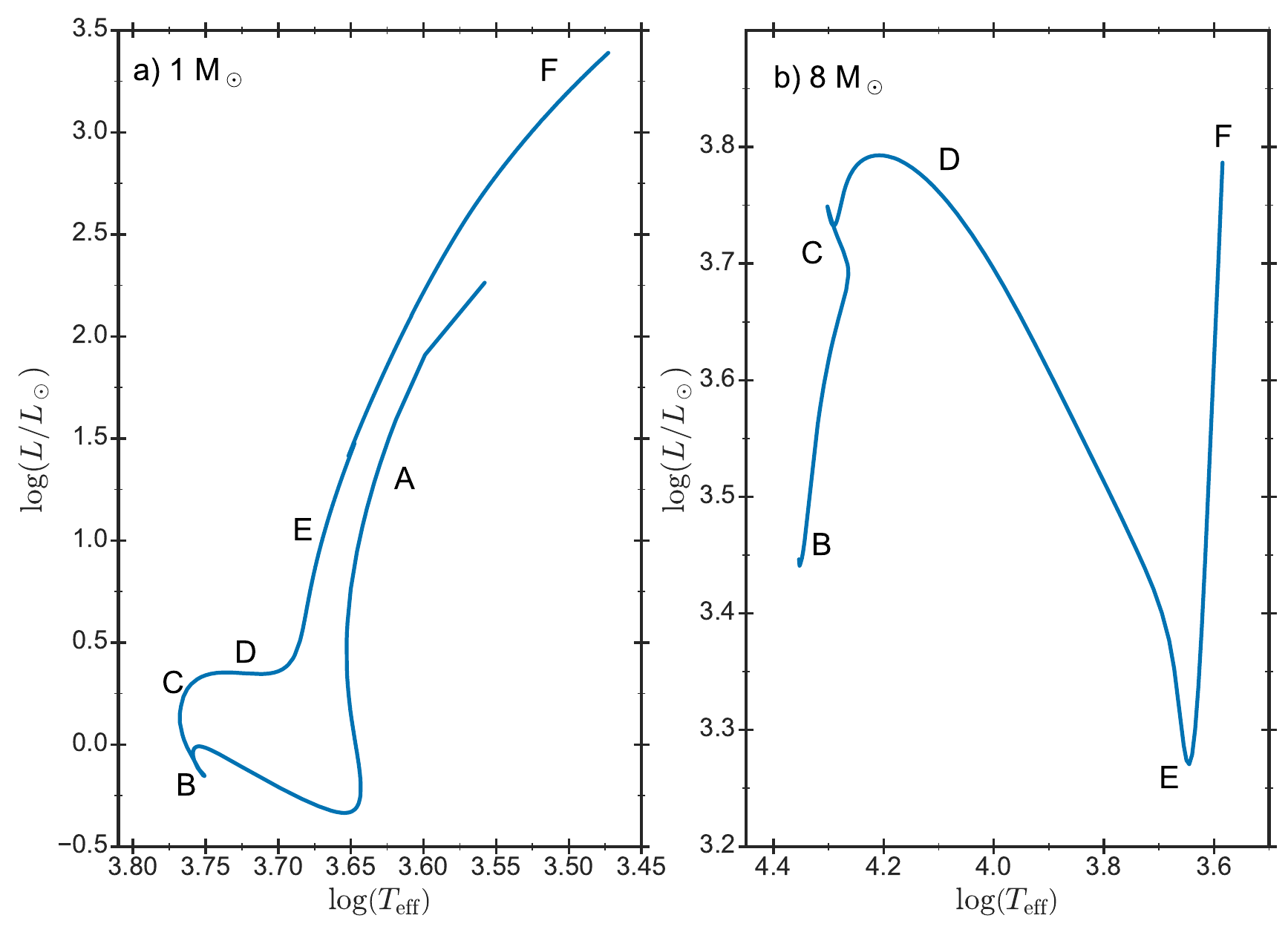}
\caption{Hertzsprung-Russell Diagram for stars at the same metallicity of those in Fig.~\ref{fig:hrd}. {\it Left panel}: Evolution of a 1~M$_\sun$ star from the pre-main sequence to the tip of the red giant branch. {\it Right panel}: Similar evolution of an 8.0~M$_\sun$ star, where the pre-main sequence has been removed for better visualization.}
\label{fig:hrd_MS}   
\end{figure}

The ultimate fate of a star depends mostly in its initial mass and chemical composition, properties that are related to the place and time where the star was born and possible interactions with the medium surrounding it. It is customary to analyse the main phases of stellar evolution following the path described by the surface luminosity, $L$, and effective temperature, $T_\mathrm{eff}$, throughout the star's lifetime. This is the so-called \textit{Hertzsprung-Russell Diagram} (HRD), and it is shown for several masses in Fig.~\ref{fig:hrd} for a given chemical composition. I will mostly focus on the evolution of stars with masses below $\sim$2.5-3.0~M$_\sun$, from the beginning of the hydrogen burning phase until helium is exhausted in the centre. Nevertheless, I will also broadly describe the evolution of more massive stars. For better guidance through the different evolutionary stages, Fig.~\ref{fig:hrd_MS} presents two HR Diagrams of different evolutionary phases for stars of the same metallicity as that of the tracks in Fig.~\ref{fig:hrd}. The left panel shows a 1~M$_\sun$ star evolving from the pre-main sequence until helium ignition, while the right panel depicts a 8~M$_\sun$ star in similar evolutionary phases.
\subsection{Pre-main sequence phase}\label{ssec:prems}
Protostars form through condensation of interstellar matter at low temperatures in hydrostatic equilibrium. Giant molecular clouds with masses large enough to undergo gravitational collapse fragment into smaller subunits due to the presence of inhomogeneities. Each of these subregions forms an hydrostatic core and accretes free-falling gas from its surroundings. Once this accretion process is complete, the protostar collapses again until hydrostatic equilibrium is restored, giving birth to a new star.

A star in this phase is fully convective, has a low temperature, a large radius and a high luminosity. It evolves almost vertically in the HRD (at roughly constant temperature) in the so-called \textit{Hayashi track} (see point `A' in the left panel of Fig.~\ref{fig:hrd_MS}), its exact location depending on the initial mass and chemical composition. Once a radiative core appears, the star leaves the Hayashi track and moves to higher effective temperatures while its convective envelope slowly retreats. The star contracts increasing its central temperature and density until fusion of hydrogen to helium becomes efficient and releases enough energy to counteract the gravitational force. At this point, the star has reached the Zero Age Main Sequence (ZAMS, point `B' in both panels of Fig.~\ref{fig:hrd_MS}).

Due to the temperature sensitivity of the reaction networks, stars of masses M~$\leq$~1.3 M$_\odot$ burn hydrogen mainly through the p-p chain (see Table~\ref{sssec:nuc,tab:chain}), the exact transition mass value depending on the initial chemical composition (it is M$\approx$1.3 M$_\odot$ at solar metallicity). Before reaching the ZAMS, some nuclear burning occurs in these stars, such as transforming deuterium into $^{3}\mathrm{He}$. While this reaction takes place and the abundance of $^{3}\mathrm{He}$ is not high enough as to complete the pp~I branch of the p-p chain, the star is forced to reach higher temperatures and densities in order to satisfy its energy needs. This higher temperature also induces the first three reactions of the CNO cycle, burning $^{12}\mathrm{C}$ into $^{14}\mathrm{N}$. A small convective core appears as a consequence of an energy generation more concentrated to the center; it only survives until the $^{12}\mathrm{C}$ abundance decreases and the amount of $^{3}\mathrm{He}$ increases enough for the p-p reactions to become more important and the energy generation to be redistributed over a larger area.
\subsection{Main sequence evolution} \label{ssec:ms}
The main sequence corresponds to the phase in which a star transforms hydrogen into helium at its centre, and it is the longest of the evolutionary phases in a star's lifetime (points `B' to `C' in Fig.~\ref{fig:hrd_MS}). Its duration is mainly controlled by the mass of the star, while in comparison the chemical composition and mixing processes play a secondary role. From homology relations (see e.g., \cite{2012sse..book.....K}), there exists a mass-luminosity relation that gives the dependence between these parameters in different evolutionary stages. For stars on the main sequence, at a given chemical composition, it is of the order of
\begin{equation}\label{ssec:ms,eq:mass_lum}
\left(\frac{L}{L_\sun}\right) \propto \left(\frac{M}{M_\sun}\right)^{3.5}\,.
\end{equation}
Recalling the nuclear time scale given by Eq.~\ref{sec:mod,eq:nuc_tim}, and replacing $L$ with the mass-luminosity relation, it is clear that the time a star can shine with nuclear burning as its energy source decreases for increasing stellar mass. To give some numbers, at solar metallicity a 1.0~M$_\sun$ star burns hydrogen for approximately 9~Gyr., while a 20~M$_\sun$ star does it for 8~Myr.

Stars more massive than $\sim$1.7~M$_\sun$ have a convective core and a radiative envelope, while their less massive counterparts have convective envelopes on top of their radiative interiors and can host either radiative or convective cores. The presence of a convective envelope has important consequences for the pulsation properties of stars, as it stochastically produces excitation of modes.

The main effect of a larger mass is a significant increase of the interior temperature, resulting in different efficiencies of the H-burning reaction networks. Depending on the resulting mechanism employed to burn hydrogen, stars are usually classified in lower main-sequence stars (masses below $\sim$1.3 M$_\odot$ where the p-p chain is the main mechanism) and upper main-sequence stars (more massive stars where the CNO cycle plays the leading role). When the CNO cycle is the dominant source of energy production, the centre of the star becomes convective due to a very high energy flux in the innermost regions. As the mass increases, convective core size also grows as a consequence of the higher temperature in the interior leading to a larger flux.

It is assumed that the convective core is homogeneously mixed and its size determines the amount of available fuel for hydrogen burning. For a star of a given mass and chemical composition, this defines the total time it will spend burning hydrogen, the size of its helium core once the hydrogen in the centre is exhausted, and the exact position of the star in the HRD. If the star had a convective core during its main-sequence evolution, its disappearance once the star reaches the \textit{turn-off} point leaves a characteristic hook-like feature in the HRD (compare point `C' in each panel of Fig.~\ref{fig:hrd_MS}). The existence of this feature in the Colour Magnitude Diagram (CMD) of clusters is used to calibrate the amount of mixing beyond the convective core and directly impacts age determinations via isochrone fitting \citep[e.g.,][]{Vandenberg:2007cq}. Despite its importance, the exact extension of the convective core is still an open problem due to the uncertainty in the `true' convective boundary definition, and the contribution of the different physical processes that mix material beyond this formal boundary (\cite[e.g.,][]{SilvaAguirre:2011jz,Gabriel:2014bu}).
\subsection{Subgiants, giants, and clump giants}\label{ssec:postms}
At the turn-off point (related to hydrogen exhaustion in the centre), hydrogen burning ceases to be a central process and becomes a shell-burning process in a layer outside of the He-rich core. At the same time, the stellar envelope expands cooling down the star and moving it to the right in the HRD. This constitutes the \textit{subgiant} phase (point `D' in Fig.~\ref{fig:hrd_MS}), where stars evolve roughly at the pace set by the Kelvin-Helmholtz time scale (Eq.~\ref{sec:mod,eq:kh_tim}).

For masses above $\sim$2.5-3.0~M$_\sun$, the core contracts since it cannot counteract the pressure exerted by the layers above it (the exact relation between core and envelope mass is given by the \textit{Sch\"onberg-Chandrasekhar} limit, \cite{Schonberg:1942hh}). A convective envelope develops due to the cooling down of the outer layers, which marks the beginning of the \textit{red giant} (RG) phase (point `E' in Fig.~\ref{fig:hrd_MS}(b)). From here onwards, the star evolves at a roughly constant temperature, burning hydrogen in a shell while increasing its luminosity and radius and further contracting the core. Eventually, the central temperature reaches values high enough as to ignite helium in the centre, marking the end of the RG phase (point `F' in Fig.~\ref{fig:hrd_MS}(b)). This occurs under non-degenerate conditions, as the central density is low enough to prevent the onset of electron degeneracy. For increasing stellar mass, the time needed to reach He-burning central temperatures is very short and the RG might even disappear.

The case of masses lower than $\sim$2.5~M$_\sun$ is slightly different from their more massive counterparts. The gas in the He-rich core is electron-degenerate, providing enough pressure to support the envelope above it and at the same time grow from the production of helium in the H-burning shell. During the subgiant phase, the cooling down of the outer layers results in an inward penetration of the already existing convective envelope, which drags partially processed nuclear material to the surface. Point `E' in Fig.~\ref{fig:hrd_MS}(a) shows the position in the HRD where maximum inward penetration of the envelope occurs, a phenomenon called the \textit{first dredge-up}. This is observationally witnessed by the change in CNO abundances due to mixing of former nuclearly processed core material dredged-up to the surface.

After this process, the convective envelope begins to retreat and the star continues its vertical ascent through the red giant branch (RGB, the portion of the HRD populated by stars evolving through the RG phase). One must keep in mind that, during the evolution up the RGB, stars lose some amount of mass through stellar winds due to their increasingly larger envelopes. Nevertheless, the mass-loss rates have not been tightly constrained by either observations or theory, and ad-hoc parametrizations are used to reproduce that phenomena such as the one given by \cite{Reimers:1975vw,Reimers:1977ts}.

\begin{figure}[h]
\centering
\sidecaption
\includegraphics[scale=0.65]{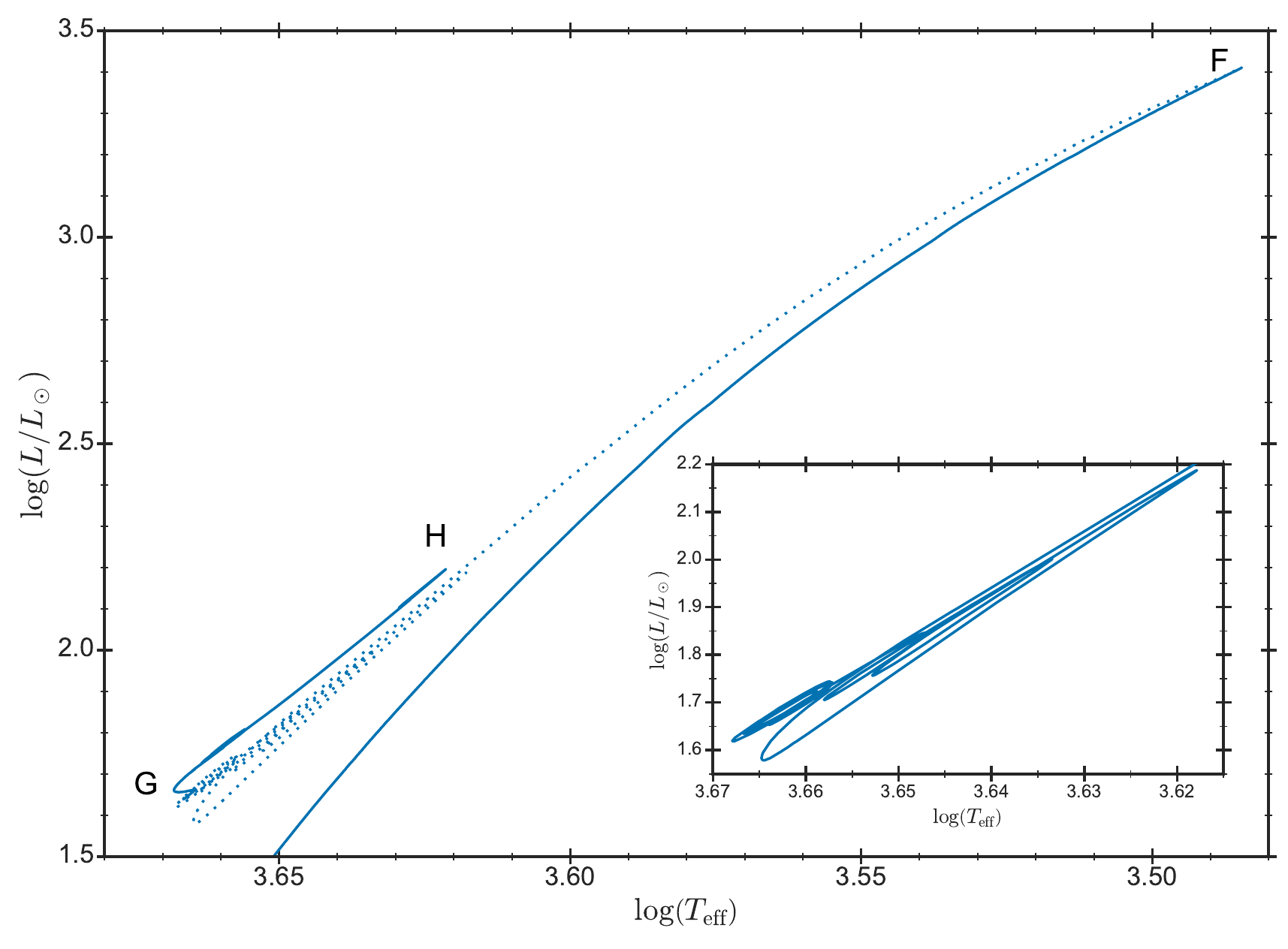}
\caption{Hertzsprung-Russell Diagram of a 1.0~M$_\sun$ star from the tip of the red giant branch until centre helium exhaustion at low metallicity. The evolution during the helium flash is marked with a dotted line in the main panel for clarity, and is also depicted in the inset (solid line).}
\label{fig:hrd_He}   
\end{figure}

Although helium ignition occurs quietly in massive stars, the case for stars below $\sim$2.5~M$_\sun$ is somewhat different. For better visualisation, Fig.~\ref{fig:hrd_He} depicts the evolution from the tip of the red giant branch until centre helium exhaustion of a 1.0~M$_\sun$ star at solar metallicity. In the late stages of the RGB, stars lose large amounts of energy in the form of neutrinos, this form of dissipation being most efficient where the stellar matter is more dense (its centre). An inversion of the thermal profile occurs in the He-core, the hottest place being a layer located off-centre within the He-rich core. When temperatures high enough to start helium burning are reached, the ignition takes place off-centre in a sort of thermonuclear runaway, called the \textit{core helium flash} (point `F' in Fig.~\ref{fig:hrd_MS}(a) and~Fig.\ref{fig:hrd_He}).

The reason for this phenomenon has to do with a property of the electron-degenerate gas, which is the decoupling of the temperature dependence from the $P-\rho$ relation (the equation of state, see Sect.~\ref{ssec:solv}). For such a case, the energy input to the medium due to the nuclear reactions when helium ignites increases the local temperature, but no compensating increase of the pressure takes place. As a consequence, the region where burning takes place does not expand to cool down the material, leading to an increase in the thermonuclear reaction rates, which leads to a further increase in the local temperature, and so forth. The thermodynamical runaway is terminated once the increasing temperature removes degeneracy and the EOS becomes temperature dependent. During this phase, the large amounts of energy produced by the ignition are used to lift up the degeneracy in the core, decreasing considerably the luminosity; in fact, secondary flashes take place increasingly closer to the centre until degeneracy has been lifted throughout the He-core, producing loops in the HRD (see inset in Fig.~\ref{fig:hrd_He}). 

The star is now able to burn helium quiescently in a convective core and hydrogen in a shell, marking the end of the RG phase. This evolutionary stage is called the \textit{horizontal branch} phase for stars in Globular Clusters (low-mass metal-poor stars, see point `G' in Fig.~\ref{fig:hrd_He}), while it is usually known as the \textit{red clump} for composite populations since their location in the HRD can be closer to the RGB. In this latter case, a secondary (and less luminous) red clump can also be present, comprising the more massive stars that started the He-burning in non-degenerate conditions (\cite{Girardi:1999jy}).

The horizontal branch or clump is the second longest evolutionary phase in the life of a star, in which helium is burned in a homogeneously mixed convective core through the 3$\alpha$ mechanism (see Sect.~\ref{sssec:nuc}), surrounded by a H-burning shell. Evolution until exhaustion of helium in the centre is shown in Fig.~\ref{fig:hrd_He} (points `G' to `H'), which for this case lasts approximately 80~Myr.
\subsection{Advanced stages of evolution}\label{ssec:adv_stag}
For the sake of completeness, I briefly mention the evolutionary phases beyond the helium burning stage. Once helium is exhausted in the core, stars with initial masses $\lesssim$~8.0~M$_\sun$ shine by helium and hydrogen shell burning. They are said to ascend the \textit{asymptotic giant branch} (AGB), a phase where extinction and re-ignition of helium leads to the occurrence of thermal pulses. During this evolutionary stage, stars undergo large amounts of mixing and complex nuclear reactions, such as slow-neutron capture and carbon burning. Large amounts of mass are lost due to dust-driven winds and large amplitude pulsations. Once hydrogen is largely exhausted in the burning shell, the remaining envelope is rapidly lost and shines due to ionisation by the bare core of the star as a \textit{planetary nebula}. This exposes the central star (a \textit{white dwarf}), which consequently evolves down the white dwarf cooling curve over a time scale of billions of years.

The value of $\sim$8.0~M$_\sun$ given above is a very loose approximation of the maximum initial mass a star should have to end its life as a white dwarf. Its ultimate fate will depend on its capability of losing enough mass by stellar winds throughout evolution to have a final mass smaller than the so-called \textit{Chandrasekhar} limiting mass. This limit gives the highest possible final core mass a star can have in order to be a stable a white dwarf, and it is usually considered to be approximately $\sim$1.45~M$_\sun$.

Naturally there are stars that have core masses far higher than the \textit{Chandrasekhar} limiting mass once they have exhausted helium in the centre. These stars will ignite carbon under non-degenerate conditions. As they go through several cycles of nuclear burning, they produce shells of heavier elements inside in a so-called \textit{onion skin} model. At last, the core consisting mostly of $^{56}\mathrm{Fe}$ (or a neighbouring nuclei) becomes dynamically unstable and core collapse sets in, resulting in a supernova explosion.
\section{Closing remarks}\label{sec:concl}
The picture of stellar structure and evolution sketched in this chapter is an overwhelmingly simplified one and I encourage the reader to consult to extensive available literature for all the interesting details about the birth, life, and fate of stars. I have reviewed the main components and assumptions entering stellar evolution calculations in different phases and regimes, highlighting the areas where large uncertainties still remain and how additional tools such as asteroseismology can help constraining these processes.

Ultimately, our knowledge of the formation and evolution of planetary systems critically depends in our understanding of stars. Several compilations of stellar tracks and isochrones including state-of-the-art macrophysics and microphysics have been computed throughout the years by different groups, and are perfectly suited for a wide range of applications in astrophysical research. Some of the most commonly used include the BaSTI isochrones (\cite{Pietrinferni:2004im,2007AJ....133..468C,Salaris:2010jj}), the Darthmouth stellar evolution database (\cite{Dotter:2008ga}), the MESA Isochrones and Stellar Tracks (\cite{Choi:2016kf}), and the PARSEC stellar tracks and isochrones (\cite{Bressan:2012bx}). These are routinely used in the characterisation of exoplanet host stars (e.g., \cite{Schlaufman:2010db,Huber:2013jb,2015MNRAS.452.2127S}), and can be combined with sophisticated Bayesian schemes to extract stellar properties given some set of observations (e.g., \cite{DaSilva:2006be,Serenelli:2013fz,SilvaAguirre:2017eh}).
\begin{acknowledgement}
The author would like to thank the organisers for their invitation to be an instructor at this Summer School, giving me the opportunity of sharing my care for stars with an outstanding group of avid young astronomers. The author also thanks J\o rgen Christensen-Dalsgaard, Achim Weiss, Santi Cassisi, and Aldo Serenelli for many stimulating discussions that helped shaping the content of this lecture.
\end{acknowledgement}
\bibliography{faial}
\end{document}